\documentclass[12pt,a4paper]{article}

\newcommand{\ka}{\kappa}

\begin{document}

\begin{flushright}
hep-th/0203233
\end{flushright}
\vspace{1.8cm}

\begin{center}
 \textbf{\Large Baryon Masses and Wilson Loops \\ for Fractional
D3-Branes on the Resolved Conifold}
\end{center}
\vspace{1.6cm}
\begin{center}
 Shijong Ryang
\end{center}

\begin{center}
\textit{Department of Physics \\ Kyoto Prefectural University of Medicine
\\ Taishogun, Kyoto 603-8334 Japan}  \par
\texttt{ryang@koto.kpu-m.ac.jp}
\end{center}
\vspace{2.8cm}
\begin{abstract}
We study the IR dynamics of the type IIB supergravity solution 
describing $N$ D3-branes and $M$ fractional D3-branes on the resolved
conifold. The baryon mass and the tension of domain wall 
in the dual gauge theory are evaluated
and compared with those for the deformed conifold. The IR behavior of
the solution for the general conifold is also discussed. 
We show that the area law behavior of the Wilson loop is attributed
to the existence of the locus in the IR where the D3-brane
charge vanishes.
\end{abstract} 
\vspace{3cm}
\begin{flushleft}
March, 2002
\end{flushleft}
\newpage
\section{Introduction}

The original AdS/CFT correspondence \cite{JM,GKP,EW} relating the
supergravity and the strongly coupled superconformal gauge theory
has been extended to the systems with less supersymmetry \cite{PS,
DM,KS,LNV,AK,KW,MP} and the non-conformal systems \cite{MD,GK,
KN,KT}. If the $N$ D3-branes are placed at the apex of a six-dimensional
cone whose base is a five-dimensional Einstein manifold $X_5$,
supersymmetry is partially broken and the type IIB string theory 
on $AdS_5 \times X_5$ is considered to be dual to 
the low energy limit of the worldvolume theory
on the D3-branes at the conifold singularity, which is a 
$\mathcal{N}=1$ superconformal gauge theory with $SU(N) \times SU(N)$
gauge group. Breaking conformal invariance can be performed by placing
the additional $M$ fractional D3-branes at the singularity.
The fractional D3-branes are thought of as D5-branes wrapped over 
collapsed 2-cycles at the singularity and modify gauge group to 
$SU(N+M) \times SU(N)$ \cite{GK,KN}. The conformal invariance is 
broken so that the relative gauge coupling runs logarithmically and the
supergravity equations are solved to leading order in $M/N$ \cite{KN}
and to all orders \cite{KT}. The solution has a naked singularity at 
small radius. The appearance of naked singularities in the gravitational
duals of the non-conformal gauge theories with reduced supersymmetry is
rather common phenomena \cite{MN,BV,JP}.

In Ref. \cite{IKS} the solution without the conical singularity has been
constructed via the deformation of the conifold where the apex is
replaced by a $S^3$. The decrease of the D3-brane charge is identified
with the repeated chain of  Seiberg-duality transformations. 
In this solution there appear no naked singularities in the IR.
At the IR end of the duality cascade the theory becomes a $\mathcal{N}=1\;
SU(M)$ gauge theory which confines with chiral symmetry breaking.
Based on this theory the spectrum of glueball masses has been calculated
\cite{CH} and the gluino condensation as well as the Wilson loop have 
been investigated \cite{LS,HKO}. The alternative approach for 
removing the naked singularity is to put the system at high temperature,
which leads to the restoration of chiral symmetry \cite{AB,BH,GH}.
On the other hand the conical singularity is removed by the resolution of
the conifold where the apex is replaced by a $S^2$ \cite{ZT,LZT}
or by the generalization of the conifold \cite{LZT}. 
In the system of $N$ regular D3-branes and $M$ fractional D3-branes on the
resolved conifold or the general conifold a naked repulson-type 
singularity emerges far in the IR and is located behind the radius where
the R-R 5-form flux vanishes. For the 
system of only the $N$ regular D3-branes
on the resolved conifold or the general conifold, the Wilson loop for a
quark anti-quark interaction is studied \cite{LZT}. 

We will further analyze the solution of $N$ D3-branes as well as $M$ 
fractional D3-branes on the resolved conifold or the general conifold to
extract IR properties for the baryon mass, the tension of domain wall 
and the tension of fractional D3-brane. The Wilson loop for the dual
non-conformal gauge theory will be shown to take the area law
behavior which is associated with the existence of the locus in the IR
where the R-R 5-form flux vanishes. 

\section{Baryon masses and domain walls}

Let us write the metric of the type IIB supergravity solution representing
$N$ regular D3-branes and $M$ fractional D3-branes on the general 
resolved conifold \cite{LZT}
\begin{equation}
ds_{10}^2 = h^{-1/2}(\rho)dx^{\mu}dx^{\mu} + h^{1/2}(\rho)ds_6^2,
\end{equation}
where the six-dimensional metric on the general resolved conifold is 
given by
\begin{equation}
ds_6^2 = \ka^{-1}(\rho)d\rho^2 + \frac{1}{9}\ka(\rho)\rho^2e_{\psi}^2
+ \frac{1}{6}\rho^2(e_{\theta_1}^2 + e_{\phi_1}^2) 
+ \frac{1}{6}(\rho^2+ 6a^2)(e_{\theta_2}^2 + e_{\phi_2}^2)
\end{equation}
with the resolution parameter $a$ and 
\begin{equation}
\ka(\rho) = \frac{1+\frac{9a^2}{\rho^2} - \frac{b^6}{\rho^6}}
{1 + \frac{6a^2}{\rho^2}},\; e_{\psi} = d\psi + \sum_{i=1}^{2}\cos
\theta_id\phi_i, \; e_{\theta_i} = d\theta_i, \; e_{\phi_i} = 
\sin \theta_id\phi_i,\; i=1, 2.   
\end{equation}
The length scale $b$ is also introduced as a generalization parameter to
smoothen the curvature singularity of the original conifold metric.
There appears an asymmetry between the two $S^2$ parts of the resolved
conifold $(a\neq 0, b=0)$, while there is no asymmetry for the general
conifold  $(a=0, b\neq 0)$. In the resolved conifold near the apex
the $S^3(\psi,\theta_1,\phi_1)$ part of the metric shrinks to zero
size and the $S^2(\theta_2,\phi_2)$ part remains finite with a radius
$a$. The $M$ D5-branes wrapped over the 2-cycle that are identified with
the fractional D3-branes serve as sources of the magnetic R-R 3-form
flux to give 
\begin{equation}
F_3 = P e_{\psi}\wedge(e_{\theta_2}\wedge e_{\phi_2} - 
e_{\theta_1}\wedge e_{\phi_1})
\label{fpe}\end{equation}
with $P \sim M\alpha'$. The type IIB equations are satisfied by the NS-NS
2-form potential and the R-R 5-form flux
\begin{eqnarray}
B_2 &=& f_1(\rho)e_{\theta_1}\wedge e_{\phi_1} +  
f_2(\rho)e_{\theta_2}\wedge e_{\phi_2}, \nonumber \\
F_5 &=& \mathcal{F} + *\mathcal{F}, \; \mathcal{F} = K(\rho)e_{\psi}
\wedge e_{\theta_1}\wedge e_{\phi_1}
\wedge e_{\theta_2}\wedge e_{\phi_2}
\label{rrf}\end{eqnarray}
together with the warp function $h$ determined from the trace of the
Einstein equation as
\begin{equation}
h' = -\frac{108g_s K}{\rho^5 \ka \Gamma},
\label{deh}\end{equation}
where $K = Q + P(f_1 - f_2), Q \sim N\alpha'^2$ and 
$\Gamma = (\rho^2 + 6a^2)/\rho^2$. Following Ref. \cite{HKO} we can
fix the normalization as 
\begin{equation}
Q = \frac{1}{4}\pi N\alpha'^2, \; P = \frac{1}{4} M \alpha'.
\label{nor}\end{equation}
For the general conifold the symmetric solution is given by
\begin{equation}
f_1 = - f_2 = \frac{1}{2}g_sP \ln (\bar{\rho}^6 - 1) + f_0, \;
K = Q + g_sP^2\ln (\bar{\rho}^6 - 1) + 2Pf_0
\label{gcs}\end{equation}
with $\bar{\rho}= \rho/b\ge 1$. In the 
following we will devote ourselves to
the case that the number of fractional D3-branes is much smaller than that
of regular D3-branes such that $g_sN \gg (g_sM)^2$ under the large
$g_sN, g_sM$ supergravity regime. In this case the 
integration constant $f_0$ can be neglected. The warp function 
$h$ obtained by substituting $K$ of (\ref{gcs}) 
into (\ref{deh}) and integrating, shows the standard conifold behavior
\begin{equation}
h = h_0 + \frac{27}{\rho^4}\left[ g_sQ + 6(g_sP)^2( 
\ln\bar{\rho} + \frac{1}{4} ) \right]
\label{gch}\end{equation}
in the large $\rho$ region for the string metric, while it gives the
short distance behavior 
\begin{equation}
h = h_0 - \frac{18g_sQ}{b^4}\ln (\bar{\rho}^2 - 1) - 
\frac{9(g_sP)^2}{b^4}\ln^2 (\bar{\rho}^2 - 1).
\end{equation}
On the other hand the asymmetric solution for the resolved conifold is
expressed as \cite{ZT}
\begin{eqnarray}
f_1 &=& \frac{3}{2}g_sP\ln (\rho^2 + 9a^2) + f_{10}, \; f_2 = \frac{1}{6}
g_sP \left(\frac{36a^2}{\rho^2} - \ln [\rho^{16}(\rho^2 + 9a^2)]\right)
+ f_{20}, \nonumber \\
K &=& Q - \frac{1}{3}g_sP^2 \left(\frac{18a^2}{\rho^2} - \ln [\rho^{8}
(\rho^2 + 9a^2)^5]\right) + P(f_{10} - f_{20}),
\label{rcs}\end{eqnarray}
where $f_{10}$ and $f_{20}$ are the constants of integration. 
Alternatively if we choose the integration constants $\rho_1, \rho_2$ 
instead of $f_{10}, f_{20}$ as the following
\begin{equation}
f_1 = \frac{3}{2}g_sP \ln \frac{\rho^2 + 9a^2}{\rho_1^2}, \;
f_2 = \frac{1}{6}g_sP \left( \frac{36a^2}{\rho^2} -
\ln \frac{\rho^{16}(\rho^2 + 9a^2)}{\rho_2^{18}}\right),
\label{ff}\end{equation}
we have 
\begin{equation}
K = Q - \frac{1}{3}g_sP^2 \left(\frac{18a^2}{\rho^2} - \frac{9}{2}\ln 
\frac{\rho^2 + 9a^2}{\rho_1^2} - \frac{1}{2}\ln 
\frac{\rho^{16}(\rho^2 + 9a^2)}{\rho_2^{18}} \right).
\label{kk}\end{equation}
The large $\rho (\rho \gg 3a)$ behavior of $h$ is also the same as
the warp function for the standard conifold
\begin{equation}
h = h_0 + \frac{27}{\rho^4}\left[ g_sQ +  \frac{3}{2}(g_sP)^2( 
\ln \frac{\rho^2}{\rho_1^2} +
\ln \frac{\rho^2}{\rho_2^2} + 1 ) \right]
\end{equation}
and in the IR region $(\rho \ll 3a)\; h$ shows the power behavior
\begin{equation}
h = h_0 + \frac{6g_sQ}{a^2\rho^2} - \frac{18(g_sP)^2}{\rho^4}.
\label{irh}\end{equation} 
Hereafter we set $h_0 = 0$ to consider the near-core region.
The warp function has a naked singularity at $\rho = \rho_h,
\rho_h^2 = 3g_sP^2a^2/Q$ and the R-R 5-form flux vanishes at 
$\rho = \rho_K, \rho_K = \sqrt{2} \rho_h$. The inequality $\rho_K >
\rho_h$ is also seen in the general conifold where the naked singularity
and the zero-charge locus $(K = 0)$ are located at $\bar{\rho_h} \approx
1 + e^{-2Q/(g_sP^2)}/2$ and $\bar{\rho_K} \approx 1 + e^{-Q/(g_sP^2)}/6$
respectively. 

Now we are ready to consider a regular D3-brane probe in the background
of $N$ D3-branes and $M$ fractional D3-branes on the general resolved
conifold. The relevant D3-brane probe action is given by
\begin{equation}
S = - T_{D3} \int d^4\sigma e^{-\phi} \sqrt{-\det (g_{ab} + 
\mathcal{F}_{ab})} + \mu_{D3} \int C_4,
\label{pa}\end{equation}
where $T_{D3}$ and $\mu_{D3}$ are the tension and basic R-R charge of the
D3-brane, and $\mathcal{F}_{ab} = B_{ab} + 2\pi\alpha'F_{ab}$.
 Here $g_{ab}$ and $B_{ab}$ are the pulls-back of the 
ten-dimensional metric and the NS-NS 2-form potential. 
Let us assume that the probe D3-brane is parallel to the source of $N$
D3-branes and $M$ fractional D3-branes and extends to the $x^1, x^2, x^3$
directions with the (infinite) volume $V_3$. Taking the Hodge dual of 
the expression $\mathcal{F}$ in (\ref{rrf}) we have
\begin{equation}
*\mathcal{F} = \frac{108K}{h^2\rho^5\ka\Gamma}d\rho
\wedge dx^0\wedge dx^1 \wedge dx^2\wedge dx^3.
\label{hof}\end{equation}
In view of $F_5 = dC_4 + B_2\wedge F_3$ together with (\ref{fpe}),
(\ref{rrf}) the $(\rho,x^0,x^1,x^2,x^3)$ component of $B_2 \wedge F_3$
vanishes so that the expression (\ref{hof}) yields
\begin{equation}
C_4 = \frac{1}{g_sh} dx^0\wedge dx^1\wedge dx^2\wedge dx^3,
\label{ch}\end{equation}
where the Eq. (\ref{deh}) has been used. The substitution of (\ref{ch})
and the background solution 
into the probe action (\ref{pa}) in the static gauge leads to
\begin{equation}
S = - V_3 \int dt\left( \frac{T_{D3}}{h} - \frac{\mu_{D3}}{g_sh} \right).
\end{equation}
This vanishing implies that the gravitational attractive force between 
a probe D3-brane and the source expressed as 
\begin{equation}
F = -T_{D3} V_3 \frac{\partial}{\partial \rho}\left( \frac{1}{h} \right)
= - T_{D3} V_3 \frac{108g_sK}{h^2\rho^5 \ka\Gamma}
\label{for}\end{equation}
cancels out against the repulsive electric force between the R-R charges
with the same sign. The gravitational attractive force (\ref{for}) 
vanishes at the zero-charge locus and diverges at the naked singularity
and becomes repulsive between the two locations. If we consider an
anti-D3-brane probe the cancellation does not happen and the total 
interaction becomes attractive. 

There is the other configuration that a probe D3-brane wraps the 3-cycle
of the general resolved conifold. The wrapped D3-brane is considered to 
play the role of a baryon vertex \cite{GK}. Since the relevant worldvolume
action has no contribution from the Wess-Zumino term, the wrapped 
D3-brane will feel the attractive force only. For the resolved conifold
a D3-brane can wrap over two kinds of 3-cycle associated with constant
$(\theta_1,\phi_1)$ or $(\theta_2,\phi_2)$. For the probe D3-brane wrapped
over $S^3(\psi,\theta_1,\phi_1)$ of the resolved conifold we substitute
the classical solution describing a collection of $N$ D3-branes and $M$
fractional D3-branes into the probe action in the static gauge to
obtain a baryon mass
\begin{equation}
M_B = \frac{8\pi^2}{9} T_{D3}\sqrt{\ka_a} \rho 
[ h\rho^4 + (6f_1)^2 ]^{1/2}
\label{mbf}\end{equation}
with $\ka_a = (\rho^2 + 9a^2)/(\rho^2 + 6a^2)$, 
which includes the nontrivial contribution from 
the NS-NS $B_2$ field. The baryon mass, that is,
the total energy of the static wrapped D3-brane is a product of 
D3-brane tension, area of $S^3(\psi,\theta_1,\phi_1)$, the gravitational
potential $\sqrt{-g_{00}}$ and so on. The variable $\rho$ represents the
position of the wrapped D3-brane, which feels an attractive radial force
$-\partial M_B/\partial \rho$ which corresponds to (\ref{for}), and 
the other forces exerted by 
$M$ fundamental strings attatched on itself. In the UV regin 
$(\rho \rightarrow \infty)$ the baryon mass is expressed as
\begin{equation}
M_B = \frac{\rho}{2\pi \alpha' g_s} \left[ \frac{1}{3}\pi g_sN_{eff}
 + (g_sM)^2\ln^2\rho \right]^{1/2}
\label{mbl}\end{equation}
with the effective D3-brane charge, $N_{eff} = N + (3/2\pi)g_sM^2
\ln \rho$. This expression may be compared with the baryon mass
$M_B = \rho N/(8\pi\alpha')$, that is linear in $N$, for a D5-brane
wrapped over $S^5$ in the $AdS_5 \times S^5$ background \cite{YI,BI}.
The attractive gravitational force due to the $N$ D3-branes is 
constant for the latter, while that for the former shows a logarithmic
increase which is dominantly attributed to the NS-NS $B_2$ field.
On the other hand in the IR region the baryon mass at the zero-charge
locus $\rho = \rho_K, \rho_K = g_sPa\sqrt{6/g_sQ}$ is shown to
scale as
\begin{equation}
M_B \sim M\frac{\rho_K}{\alpha'}. 
\label{mbs}\end{equation}     
Compared to the baryon mass (\ref{mbf}) we have the mass of
wrapped D3-brane
\begin{equation}
M_{D3} = \frac{8\pi^2}{9} T_{D3}\sqrt{\ka_a} (h\rho^4)^{1/4} 
[ h\rho^4 + (6f_1)^2 ]^{1/2},
\label{mdh}\end{equation}
which does not include the gravitational potential $\sqrt{-g_{00}}$.
For the conifold case $(a = 0)$ with no fractional D3-branes this mass
is independent of $\rho$ and leads through the normalization (\ref{nor})
to the conformal dimension $\Delta = 3N/4$ of the corresponding 
baryonic vertex operator in the dual conformal field theory \cite{GK}.
From (\ref{mdh}) we note that the mass of
wrapped D3-brane localized at a constant $(\theta_2,\phi_2)$ vanishes
at the naked singularity $\rho = \rho_h$. Although there is a repulsive
component of the radial force $-\partial M_B/\partial \rho$ in the
region $\rho_h < \rho < \rho_K$, the $\rho$-dependent attractive force
prevails over it. The wrapped D3-brane is always pulled toward the apex
and then becomes massless at the 
naked singularity. This behavior is compared
with the $M = 0$ conifold case where the wrapped D3-brane with the 
constant mass is pulled by a constant attractive force because of the
$\rho$-independence of $h\rho^4$ in (\ref{mbf}). Here we consider the
other 3-cycle defined by the subspace at a constant value of
$(\theta_1,\phi_1)$ in the metric. The D3-brane wraps over $(\psi,
\theta_2,\phi_2)$ coordinates to yield a baryon mass
\begin{equation}
M_B = \frac{8\pi^2}{9} T_{D3}\sqrt{\ka_a} \rho 
[ h\rho^4 + (6f_2)^2 ]^{1/2},
\end{equation}
whose large $\rho$ behavior is the same as (\ref{mbl}). At the IR point
specified by $\rho = \rho_K$ it takes
\begin{equation}
M_B \sim \frac{a}{\sqrt{g_sQ}\alpha'^2} \left( Q^2 + 
\frac{g_s^2P^4}{2} \right)^{1/2},
\end{equation}
which scales as 
\begin{equation}
M_B \sim M \frac{\sqrt{g_sN}a}{g_sM\alpha'}\left( 1 + 
\frac{(g_sM)^4}{32\pi^2 (g_sN)^2} \right)^{1/2}.
\label{mbt}\end{equation}
This result shows a factor of $M$ times the 't Hooft scalings $g_sN$ and
$g_sM$. For small number of fractional D3-branes the baryon mass is
approximately expressed as
\begin{equation}
M_B \sim M \frac{a^2}{\rho_K \alpha'}
\label{snf}\end{equation}
because the NS-NS 2-form potential yields the dominant contribution.

In the general conifold the baryon mass for the D3-brane wrapped over
$S^3(\psi,\theta_1,\phi_1)$ is also estimated as 
\begin{equation}
M_B \sim \frac{Qb}{g_sP\alpha'^2} e^{-Q/2g_sP^2},
\end{equation}
where the exponential damping arises from the factor $\sqrt{\ka_b}$ with
$\ka_b = (\rho^6 - b^6)/\rho^6$ and the NS-NS flux contribution, 
$(6f_1)^2$ has the same order as the gravitational contribution through 
the warp factor, $h\rho_K^4$. This expression also shows the expected 
scaling
\begin{equation}
M_B \sim M \frac{g_sNb}{(g_sM)^2\alpha'}e^{-2\pi g_sN/(g_sM)^2},
\label{mex}\end{equation}
which has some resemblance to (\ref{mbt}) up to the exponential factor.

For comparison let us return to the standard conifold.  The warp function
is given by (\ref{gch}) with $\ln (\rho/b)$ replaced by $\ln(\rho/\rho_0)$
where $\rho_0$ is introduced as a constant of integration. The NS-NS
2-form potential and the charge function $K(\rho)$ are provided by 
(\ref{ff}), (\ref{kk}) with $a = 0$ and $\rho_1 = \rho_2 \equiv \rho_0$.
The baryon mass is also evaluated at the zero-charge locus 
$\rho_K = e^{-Q/6g_sP^2}\rho_0$ as
\begin{equation}
M_B \sim \frac{\rho_0}{g_sP\alpha'^2} e^{-Q/6g_sP^2} \left( Q^2 + 
\frac{g_s^2(3P)^4}{2} \right)^{1/2},
\end{equation}
whose leading part scales as 
\begin{equation}
M_B \sim M \frac{g_sN\rho_0}{(g_sM)^2\alpha'} e^{-2\pi g_sN/3(g_sM)^2}.
\label{mbc}\end{equation}
Although this expression is similar to (\ref{mex}), the exponential 
damping factor is here produced by 
the linear factor $\rho_K$ and the NS-NS
flux gives the main contribution. The dimensions of baryon masses are 
specified by the resolution parameter $a/\alpha'$ in (\ref{mbs}),
(\ref{mbt}) for the resolved conifold and similarly by the generalization
parameter $b/\alpha'$ in (\ref{mex}) for the general conifold. They are
contrasted to the factor $\rho_0/\alpha'$ in (\ref{mbc}) for the conifold.
The factor  $\rho_0$ is provided as an integration constant, whereas the
emergences of $a, b$ in the baryon mass at the IR locus indicate
the dimensional transmutation.

In the deformed conifold \cite{IKS,HKO} 
the radial parameter $\tau$ related
with the radial coordinate $\rho$ as $\rho^3 \sim \epsilon^2 e^{\tau}$
can be made zero, where $\rho$ approaches to $\epsilon^{2/3}$ and the 
$S^2$ part of the metric shrinks to zero while the radius of $S^3$ part
approaches to finite value that is of order $\sqrt{g_sM}$. The baryon mass
for the D3-brane wrapped over the finite $S^3$ part is estimated as
$M_B \sim mM$ where $m \sim \epsilon^{2/3}/\alpha'$ is a parameter setting
the 4-dimensional mass scale. This estimation for the deformed conifold 
is similar to that for the resolved conifold although there is some
difference of situation. The expression (\ref{mbs}) is produced for the
D3-brane wrapped over the shrinking $S^3(\psi,\theta_1,\phi_1)$ whereas
(\ref{snf}) is for the D3-brane over the $S^3(\psi,\theta_2,\phi_2)$ 
whose $S^2(\theta_2,\phi_2)$ part has a finite radius near the apex.
The parameter $\rho_K$ that is proportional to the resolution parameter
$a$ makes a role of setting the 4-dimensional mass scale, while the 
extra parameter $b$  makes the same role in the general conifold.
For the deformed conifold  the theory at $\tau = 0$ provides the 
 $SU(M)$ gauge theory with no matter through the chain
of Seiberg dualites. Analogously for the general or resolved conifold 
the far IR locus is so specified by the zero-charge locus that the
regular D3-branes disappear and only the $M$ fractional D3-branes
remain. Hence the supergravity theory at the zero-charge locus 
for the general or resolved conifold is
considered to be dual to the $SU(M)$ gauge theory.
The obtained baryon masses are shown to be linear in $M$ as 
expected from this view point.  

Here we also analyze the domain walls that are regarded as D5-branes
wapping the $S^3$ and spanning the $x^0, x^1, x^2$ directions for the
resolved conifold. Far in the IR the effective $SU(M)$ gauge theory 
has the inequivalent vacuum states between which the domain wall 
interpolates. The tension of a D5-brane wrapped over 
$S^3(\psi,\theta_1,\phi_1)$ as well as $x^0, x^1, x^2$ coordinates
at the IR locus $\rho = \rho_K$ is estimated as
\begin{equation}
T_{DW} = \frac{8\pi^2}{9} T_{D5}\sqrt{\frac{\ka_a}{h}}\rho_K
[h\rho_K^4 + (6f_1)^2 ]^{1/2} \sim \frac{\rho_K^3}{g_s\alpha'^3},
\end{equation}
whereas the tension for the wrapping over $S^3(\psi,\theta_2,\phi_2)$
is mainly given by
\begin{equation}
T_{DW} \sim \frac{\rho_K a^2}{g_s\alpha'^3}.
\end{equation}
The former behavior is compared with $T_{DW} \sim (\epsilon^{2/3}/
\alpha')^3/g_s$ for the deformed conifold \cite{IKS,HKO}. Through the
wrappings of a D5-brane and a D3-brane over $S^3(\psi,\theta_1,
\phi_1)$ for the resolved conifold the tension of  domain wall and 
the baryon mass, where the gravitational contribution is the same
order as the $B_2$ contribution, show the same forms as those for
the deformed conifold where the $B_2$ contribution is not taken
into account. Hence $\rho_K/\alpha'$ corresponds to $m \sim 
\epsilon^{2/3}/\alpha'$. On the other hand in the wrapping over
$S^3(\psi,\theta_2,\phi_2)$ for the resolved conifold the tension of
domain wall and the baryon mass where the $B_2$ contribution is more
dominant than the gravitational one, show the similar forms to 
those for the deformed conifold, but with some slight difference. 

\section{Tensions of fractional D-branes}

We are concerned with the tensions of D5-brane wrapped over 
$S^2(\theta_i,\phi_i), i = 1, 2.$  They can be also evaluated and shown
not to vanish at the naked singularity. This non-vanishing is due to
the nontrivial contribution of NS-NS $B_2$ field.
The tension of wrapped D5-brane at the zero-charge locus
scales as $T_{FD3} \sim T_{D3} g_sM$
for a two-sphere $S^2(\theta_1,\phi_1)$ inside  
$S^3(\psi,\theta_1,\phi_1).$  For $S^2(\theta_2,\phi_2)$ it scales as
$T_{FD3} \sim T_{D3} g_sN/g_sM$ which is represented by 
$T_{D3} g_sM(a/\rho_K)^2.$ Both tensions are independent of the
resolution parameter. Since the $S^2(\theta_1,\phi_1)$ shrinks to
zero size near the apex, the former is rather considered to be the tension
of fractional D3-brane because a fractional D3-brane is regarded 
as a D5-brane wrapped over a vanishing 2-cycle.

Let us turn to the type II supergravity solution corresponding to 
$M$ wrapped D$(p+2)$-branes ( fractional D$p$-branes ) and $N$ regular
D$p$-branes on a $(9-p)$-dimensional conical transverse space whose 
base is an $(8-p)$-dimensional Einstein manifold $X_{8-p}$ \cite{HK}.
The dilaton, NS-NS 2-form potential and the warp function are expressed
by using a radial coordinate $r$ here as
\begin{eqnarray}
e^{4\phi} &=& h(r)^{3-p}, \; B_2 = f\omega_2 =
P\frac{r^{p-3}}{p-3} \omega_2, \nonumber \\
h(r) &=& \left(\frac{q}{r}\right)^{7-p} - \frac{P^2}{(3-p)(10-2p)
r^{10-2p}}
\end{eqnarray}
for $p \neq 3, 5,$ where $P \sim g_sM\alpha'^{(5-p)/2}$ and
$q^{7-p} \sim g_sN\alpha'^{(7-p)/2}$. The $X_{8-p}$ is assumed to have
a 2-cycle associated with a harmonic 2-form $\omega_2$ 
and the fractional Dp-brane is the D(p+2)-brane wrapped over 
the 2-cycle.  The $\omega_2$ may be normalized such 
that $\omega_2 \wedge *_{8-p} \omega_2$ is the volume form on $X_{8-p}$.
Hereafter we restrict ourselves to $p < 3$ where the naked singularity
appears at $r = r_h$ with $r_h^{3-p} = P^2/((3-p)(10-2p)q^{7-p})$.
We estimate the IR scale $r_I$ here simply from $h'(r)=0$ as
\begin{equation}
r_I^{3-p} = \frac{P^2}{(3-p)(7-p)q^{7-p}}.
\end{equation}
From $r_I^{3-p} = ((10-2p)/(7-p))r_h^{3-p}$ we note that
$r_I > r_h$ for $p < 3$. If we assume that the 2-cycle is a $S^2$,
then the tension of wrapped D$(p+2)$-brane is given by
\begin{equation}
T_{FDp} = 4\pi T_{D(p+2)}h^{(p-3)/4} ( \alpha hr^4 + f^2 )^{1/2},
\label{tfd}\end{equation}
where there is the contribution of dilaton and 
$\alpha$ is some constant that 
depends on $p$ through the embedding of
the 2-cycle in $X_{8-p}$. As the tension of fractional D$p$-brane 
in the IR we evaluate (\ref{tfd}) at $r = r_I$ as
\begin{equation}
T_{FDp} \sim \frac{T_{Dp}}{\alpha'} P^{(p-1)/2} r_I^{(p-3)^2/2},
\end{equation}
where the contribution of NS-NS flux is the same order as 
the gravitational contribution. Its scaling is expressed as
\begin{equation}
T_{FDp} \sim T_{Dp} (g_sM)^{(p-1)/2}\left(\frac{r_I}{\sqrt{\alpha'}}
\right)^{(p-3)^2/2},
\end{equation}
which is compared with the tension of fractional D3-brane
for the resolved conifold that is linear in $M$. 

\section{Area law behaviors of Wilson loops}

We investigate the Wilson loop along two space directions in the solution
of $N$ D3-branes and $M$ fractional D3-branes on the general resolved
conifold in the Euclidean metric. Since the string is stretched in the
radial direction, the relevant action in the static gauge for the spatial
Wilson loop is given by 
\begin{equation}
S = T T_F  \int dx \sqrt{h^{-1} + \ka^{-1} (\partial_x \rho)^2},
\label{rea}\end{equation}
where $T_F = 1/2\pi \alpha'$ and there is no NS-NS flux contribution for
this configuration. From the prescription in Refs. \cite{JMA,RY},
the energy of the string configuration is given by
\begin{equation}
E = \frac{S}{T} = T_F \int_{\rho_0}^{\infty} \frac{d\rho}
{\sqrt{\ka(1 - c_0^2h)}},
\end{equation}
since the static solution is represented by $h^{-1}/\sqrt{h^{-1}
+ \ka^{-1}(\partial_x\rho)^2} = c_0$.
Owing to $c_0^2 = 1/h(y_0)$ it is further rewritten by
\begin{equation}
E = \frac{T_Fa}{2} \int_{y_0}^{\infty} \frac{dy}{\sqrt{y\ka}} 
\frac{\sqrt{h(y_0)}}
{\sqrt{h(y_0) - h(y)}},
\label{set}\end{equation}
where $y = \rho^2/a^2$ for the resolved conifold. For the general 
conifold  $y = \rho^2/b^2$ and $a$ in (\ref{set}) is replaced by $b$.
The $y_0 = \rho_0^2/a^2$ or $\rho_0^2/b^2$ is
the minimal value of $y$. The distance $l$ between a quark 
and an anti-quark is expressed in terms of a parameter $y_0$ as 
\begin{equation}
\frac{l}{2} = \frac{a}{2}\int_{y_0}^{\infty} \frac{dy}{\sqrt{y\ka}}
\frac{h(y)}{\sqrt{h(y_0) - h(y)}}.
\label{dq}\end{equation}
We focus on the long distance behavior of the potential energy 
between a quark and an anti-quark. The main contribution to the integral
in (\ref{dq}) is provided by the region near $y = y_0$.
Since the factor $(h(y_0) - h(y))^{1/2}$ is approximately expanded by
$(-(y-y_0)h'(y_0) - (y-y_0)^2h''(y_0)/2 - \cdots )^{1/2}$ in this 
region, we take the limit $y_0 \rightarrow y_K =
\rho_K^2/a^2$ or $\rho_K^2/b^2$ for the distance $l$ that is a
function of $y_0$. Then we obtain the logarithmic divergence of $l$
because $h'(y_K)$ vanishes. In this limit the main contribution to
the integral in (\ref{set}) also comes from the same region.
Combining (\ref{dq}) with (\ref{set}) we have 
\begin{equation}
E = \frac{1}{4\pi \alpha'\sqrt{h(y_0)}}l + \frac{T_Fa}{2\sqrt{h(y_0)}}
\int_{y_0}^{\infty}\frac{dy}{\sqrt{y\ka}} \sqrt{h(y_0) - h(y)}.
\end{equation}
Hence the first term is leading in the limit 
$y_0 \rightarrow y_K$ and gives rise to the
area law behavior of the Wilson loop. The tension of the confining
string is thus derived as 
\begin{equation}
T_{cs} = \frac{1}{4\pi \alpha' \sqrt{h(y_K)}}.
\end{equation}

For the resolved conifold the tension is obtained by
\begin{equation}
T_{cs} \sim \frac{1}{g_sM} \frac{(g_sM)^2a^2}{g_sN \alpha'^2},
\end{equation}
which is expressed as 
\begin{equation}
T_{cs} \sim \frac{1}{g_sM} \left(\frac{\rho_K}{\alpha'}\right)^2.
\end{equation}
This result takes the same form as the tension 
$(\epsilon^{2/3}/\alpha')^2/g_sM$ for the deformed conifold 
\cite{IKS,HKO}. We have again observed the correspondence 
between $\rho_K/\alpha'$ and $m$. For the general conifold the
tension shows the similar behavior
\begin{equation}
T_{cs} \sim \frac{1}{g_sM} \frac{(g_sM)^2b^2}{g_sN \alpha'^2}.
\end{equation}
Indeed there is a difference between the warp function in the IR region
characterized by the logarithmic  function for the general conifold and
the warp function showing the power behavior for the resolved conifold,
but the resulting tensions of the confining string turn out to be the
same form where $a$ and $b$ give the scales where non-perturbative
effects become relevant in the dual gauge theory. In order to extract
the large $l$ behavior we need to take the limit $y_0 \rightarrow
y_K$ and observe that the vanishing of derivative of the warp function
at the zero-charge locus plays an important role to obtain the area
law behavior for the Wilson loop.

\section{Conclusions}

Although there appears the naked singularity in the resolved conifold
solution or the general conifold solution of $N$ D3-branes and $M$
fractional D3-branes, we have demonstrated that the IR solution located
at the zero-charge locus of the R-R 5-form field, behind which the
naked singularity exists, gives the interesting non-perturbative 
dynamics of its dual $SU(M)$ gauge theory in the confining phase.
For the resolved conifold the baryon mass as well as the tension of
domain wall at the zero-charge locus are characterized by the resolution
parameter through the dimensional transmutation in a similar way that
those at the tip $\tau = 0$ for the deformed 
conifold are specified by the deformation 
parameter. We have evaluated them using the resolved conifold metric
itself, whereas the rescaled one of the deformed conifold metric has
been used where the scaling is performed in the transverse part of the
metric so that its effect is reflected to
produce the 4-dimensional mass scale $m$
in the longitudinal part. We have assumed that the resolved conifold
solution should not be able to continue to the region where the 
D3-brane charge is negative, and focused on the IR behavior at the
zero-charge locus. Since the resolved conifold is constructed from the
conifold by allowing an asymmetry between the two $S^2$ parts in the 
metric we have seen that there are two types of probe configurations
associated with the baryon and 
the domain wall according to which $S^2$ is included
into the wrapped $S^3$. Specially when the D3-brane or the D5-brane
wraps $S^3$ including the $S^2$ with a finite radius near the apex,
the $B_2$ contribution to the baryon mass and the tension of domain
wall is more dominant than the gravitational one. We have also observed
that the $B_2$ contribution prevents the tension of
fractional D3-brane from vanishing at the naked singularity 
in the resolved conifold and this tension becomes linear in $M$
at the zero-charge locus.

We have shown the area law behavior of the Wilson loop for the
$N$ D3-branes and $M$ fractional D3-branes on the resolved or 
general conifold, and demonstrated that the tension of the confining
string takes the same behavior as that for the deformed conifold.
The suitable correspondence between the 
$\rho_K/\alpha'$ for the resolved conifold and the mass scale
$m$ for the deformed conifold has been
presented in the baryon mass and the tensions of the domain wall
and the confining string. In Ref. \cite{LZT} the behavior of Wilson
loop was analyzed for only the $N$ regular D3-branes on the 
resolved or general conifold. Since there is no
zero-charge locus, the distance between a quark and an anti-quark
does not exhibit the logarithmic increase. The existence of the 
zero-charge locus owing to the additional $M$ fractional D3-branes
yields the logarithmic increase of the distance
and hence is important for the 
area law behavior of the Wilson loop. For the other non-conformal
supergravity solutions with less supersymmetry, it would be
interesting to study the NS-NS two-form contribution to the
tensions of various D-branes wrapped over some cycle associated with
the NS-NS flux and ask how the non-conformal property provided by
the fractional D-branes is connected with the non-perturbative
property in the IR.

\end{document}